\documentstyle[ltwol,axodraw]{article}

\def\Journal#1#2#3#4{{#1} {\bf #2}, #3 (#4)}

\def\PLB{{\em Phys. Lett.}  B}

\def\PRD{{\em Phys. Rev.} D}

\def\be{\begin{equation}}
\def\ee{\end{equation}}
\def\bea{\begin{eqnarray}}
\def\eea{\end{eqnarray}}
\begin{document}

\title{On A New Class of Models for  Soft CP Violation}

\author{David Bowser-Chao$^{(1)}$,
Darwin Chang$^{(2,3)}$,
and
Wai-Yee Keung$^{(1)}$ }
\address{$^{(1)}$Physics Department, University of Illinois at Chicago,
IL 60607-7059, USA\\
$^{(2)}$NCST and Physics Department,
National Tsing-Hua University, Hsinchu 30043, Taiwan, R.O.C.\\
$^{(3)}$Institute of Physics, Academia Sinica, Taipei, R.O.C.\\}
\vspace{.5cm}
\twocolumn[\maketitle\abstracts{ 
We elaborate on a new class of models proposed recently by us 
and compare with another class proposed by Georgi and Glashow(GG). The
models can be roughly classified as the righted-handed (or
left-handed) models in our (or GG's) case.  Both classes of models use
softly broken CP symmetry to suppress tree level KM phase as well
as the strong CP phase.  The measured value of the CP-violating parameter
$\epsilon$ are accounted for by employing a new heavy sector of
scalars and vectorial fermions.  The models can be milliweak or
superweak in nature depending on the scale of the heavy sector.  We
review the phenomenology of the right-handed models and compare with
the left-handed models.
\vspace{.3cm}
\begin{center}
{(Contribution to International Conference on High Energy Physics, 1998,
Triumf, presented by DC)}
\end{center}
}]
%
\section*{Introduction}

Recently, two new classes of models~\cite{bck,gg} of soft CP violation
have
been proposed as alternatives to the standard KM model~\cite{km}.  The
models aim at
reproducing the attractive characters of KM model and at the same time
solve the long-standing strong CP problem.

These models share the features of imposing soft (or spontaneous) CP
violation in order to avoid tree level KM phase as well as the tree level
strong CP $\theta$ phase.  
To account for the measured value of the CP-violating parameter $\epsilon$
of the neutral kaon system, the models employ a new heavy sector of
scalars and vectorial fermions.  The first class of models, that we shall
roughly classified as the right-handed (RH) models~\cite{bck}, uses a
heavy sector
that couples only to right-handed down type quarks, which are
$SU_L(2)$ singlets.
Instead, the alternative left-handed (LH) models~\cite{gg} uses new
particles  
that couple only to the ordinary left-handed
quarks, which are $SU_L(2)$ doublet.  
In this report, we review the phenomenology of the
RH models and make comments on the LH models as we go for comparison.  The
more extensive presentation is under preparation~\cite{bck2}.  Note
that the ideas in these directions were presented by Barr~\cite{barr}
and his collaborators some time ago.

The heavy sector of a typical model for our purposes requires two
additional Higgs singlets, $h_\alpha (\alpha=1,2)$ of charge $q_h$ and a
vectorial pair of heavy fermions, $Q_{L,R}$, of electromagnetic charge
$-{1\over3}+q_h$.  One can also choose to have two pairs of fermions and
only one heavy Higgs singlet.  In addition, One can choose to assign the
scalars or the fermions to be carrying the color such that together they
will couple to the
right-handed down type quarks, $d_{Ri}$.  In case of neutral, colorless
fermions (the neutrinos), it is not even necessary to have vectorial
pairs.  
One can also use the freedom in
choosing charge $q_h$ to avoids fractionally charged hadrons.  Most of the
phenomenology mention below are more or less independent of the choice of
$q_h$ and color.  Relevant new terms in the Lagrangian are:
\begin{eqnarray}
{\cal L}_{h_i} &=
  \left[
  (g \lambda_{i\alpha} \bar Q_L d_{iR} h_\alpha
+  M_Q \bar Q_L Q_R) + \hbox{h.c.} \right] 
\nonumber\\
&  - (m^2)_{\alpha\beta} {h_\alpha}^{\dag} h_\beta
  - \kappa_{\alpha\beta}
      (\phi^{\dag} \phi-|\langle\phi\rangle|^2) \,h_\alpha^{\dag} h_\beta  
\nonumber\\
&  - \kappa'_{\alpha\beta\gamma\delta}
       h_\alpha^{\dag} h_\beta h_\gamma^{\dag} h_\delta  
\; ,\label{eq:lagrangian}
\end{eqnarray}
where $\phi$ is the Standard Model Higgs doublet.  
The soft breaking of CP symmetry implies a special basis where $\lambda,
\kappa, \kappa'$ and the SM Yukawa couplings are real.  
If fermions carry color, we also require (see below) that dim-3 couplings,
$M_Q$, are real to avoid tree level contribution to $\theta$.
This leaves only a single CP violating parameter:
Im$(m^2)_{12}$.  
We can diagonalize $(m^2)_{\alpha\beta}$ by a unitary
matrix $U_{\alpha i}$ which in general is complex: $h_\alpha =
U_{\alpha i} H_i$, with $H_i$ the mass eigenstates. 
The quark-Higgs interaction in the mass eigenstate basis becomes
\begin{equation}
{\cal L}_{QqH}=g\sum_{q=d,s,b}\xi_{qj}
	(\bar Q_L q_R)H^-_j \ +\ \hbox{h.c.} \ ,
\end{equation}
where $H_i = U_{\alpha i}^* h_\alpha$ is the mass eigenstates, 
$\xi_{qj} \equiv \lambda_{q\alpha} U_{\alpha j}$. The 
Yukawa couplings $\xi_{qj}$ are defined relative to the gauge coupling
$g$ of $SU_L(2)$.
The rephasing-invariant measure of CP violation are
\begin{equation}
{\cal A}^{qq'}_{ij} = \lambda_{q\alpha}\lambda_{q'\beta}U_{\beta i}
U_{\alpha j}^{*} = \xi_{qi}^{*} \xi_{q'j}
\end{equation}
with $(q,q'=d,s,b)$ and $i,j= 1,2$.
For flavor conserving amplitudes like EDM, we define the counterpart
${\cal B}_{ij} = \kappa_{\alpha \beta}U_{\beta i} U_{\alpha j}^{*}$.
${\cal A}^{qq}_{ij}$ as well as ${\cal B}_{ij}$ are Hermitian in
indices $i.j$. Thus CP is broken only in the off-diagonal terms.  
As a result, they
contribute to CP violation only when both the light and heavy charged
Higgs are involved in a diagram at the lowest order.  
It is also possible to avoid the CP violating part of the coupling
$\kappa$ if one chooses to have
only one heavy Higgs singlet and breaks CP symmetry in the dimension-3
heavy fermion mixing terms instead as long as these heavy fermions are
colorless.  The contribution related to 
parameter $\kappa'$ in the last term of (1) 
ocurs only at the higher loop level and generally
can be ignored.

Before continuing, we like to emphasize again that if the new fermions carry
color it is necessary to impose CP symmetry on their bare masses also in
order to avoid tree level strong CP problem.  When there are more than one
pairs of vectorial fermions, one can potentially make the mass matrix
Hermitian, however that would require some additional symmetry to be 
implemented.

\section*{Constraint from $\epsilon$}

With CP broken only softly, the CKM matrix is real at tree level.  Leading
contribution to the CP violating parameter $\epsilon$ is due to the box
diagram with only heavy particles in loop.  
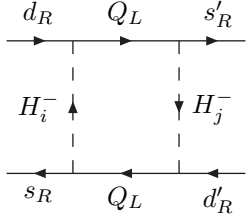
\begin{center}
\begin{picture}(110,90)(0,-20)
\ArrowLine(0,50)(25,50)   \Text(12,58)[b]{$d_R$}
\ArrowLine(25,50)(65,50)  \Text(45,58)[b]{$Q_L$}
\ArrowLine(65,50)(90,50)  \Text(80,58)[b]{$s_R'$}
\ArrowLine(25,0)(0,0)     \Text(12,-5)[t]{$s_R$}
\ArrowLine(65,0)(25,0)    \Text(45,-5)[t]{$Q_L$}
\ArrowLine(90,0)(65,0)    \Text(80,-5)[t]{$d'_R$}
\DashArrowLine(25,0)(25,50){5} \Text(20,25)[r]{$H_i^-$}
\DashArrowLine(65,50)(65,0){5} \Text(70,25)[l]{$H_j^-$}
\end{picture}
\end{center}
We evaluate all possible diagrams and match them with the low energy
effective Hamiltonian which has the form
\begin{equation}
{\cal H}^{\Delta S=2} =
{G_F^2 m_W^2 \over 16\pi^2}
\sum_{I=R,L}C^I_{\Delta S=2}(\mu) O^I_{\Delta S=2}(\mu) \ , \ 
\end{equation}
\begin{equation}
{\rm with\ } \quad\quad O^{R,L}_{\Delta S=2} =
\bar s \gamma_\mu (1\pm\gamma_5) d \,
\bar s \gamma^\mu (1\pm\gamma_5) d  \ . \end{equation}
The $W^{\pm}$  diagrams yield a purely real Wilson coefficient
$C^L_{\Delta S=2}(\mu)$; CP violation is due
solely to the operator $O^R_{\Delta S=2}$ rather than $O^L_{\Delta
S=2}$, in contrast to the KM model, because the  complex coefficient
$C^R_{\Delta S=2}(\mu)$ is generated by the charged Higgs.
At the scale $\mu=M_Q$, we have
$$
  C^R_{\Delta S=2}(M_Q) =
  2 \xi_{d1} \xi_{s1}^* \xi_{d2} \xi_{s2}^{*}
	\frac{m_W^2}{M_Q^2} \frac{f(x_2)-f(x_1)}{x_2-x_1} $$
\begin{equation}
  \quad +\sum_{i=1,2} (\xi_{di}\xi_{si}^{*})^2 \frac{m_W^2}{M_Q^2}\,
{df \over dx}(x_i)
\ , \end{equation}
with $f(h)=(1-h)^{-2}(1+2h+h^2+h^2\ln h)$. 

The real part of the diagram
can contribute to part of the $K_L-K_s$ mass difference while the
imaginary gives rise to $\epsilon$.  This is analyzed in detail in
Ref.~\cite{bck}.  For illustration here, we can take the (decoupling) limit
$m_2 \gg m_1$ and assume $m_1 = M_Q$ for simplicity.  Demanding that the
imaginary part of the box diagram gives enough
contribution to $\epsilon$ while the corresponding real part gives just a
fraction $\cal F$ of the mass difference $\Delta m_K$, we obtain~\cite{bck}
the constraints, 
\begin{eqnarray}
\hbox{Im} \left({\cal A}_{sd} / (0.058)^2 \right)^2 R_Q^2 &=1      \ ,
\\
\quad
\hbox{Re} \left({\cal A}_{sd} / (0.058)^2 \right)^2 R_Q^2 &=156 {\cal F}\ ;
\label{eq:dmineq}
\end{eqnarray}
where $R_Q = 300 \hbox{ GeV}/M_Q$. The reasonable constraint $|{\cal F}|
< 1$ can be easily satisfied.
It is important to emphasize that, in RH models, the heavy particle box
diagrams induce a right-handed four fermion operator, contrary to the LH
models and the KM model in which the leading CP violating operators are
left-handed.

\section*{Constraints from $(\epsilon'/\epsilon)$}

The leading contribution to the direct CP violating parameter $\epsilon'$
is due to the gluonic penguin diagrams with only the heavy particles in
the loop.  
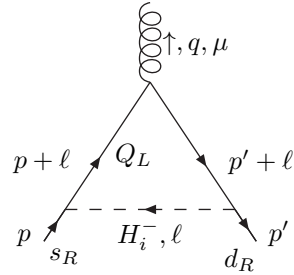
\begin{center}
\begin{picture}(100,115)(0,-15)
\ArrowLine(0,0)(40,60)
\ArrowLine(0,0)(8,12)          \Text(-5,0)[rb]{$p$} \Text(2,-3)[lt]{$s_R$}
\ArrowLine(40,60)(80,0)
\ArrowLine(72,12)(80,0)        \Text(85,0)[lb]{$p'$}\Text(80,-3)[rt]{$d_R$}
\Text(10,30)[r]{$p+\ell$}      \Text(27,30)[lb]{$Q_L$}
\Text(70,30)[l]{$p'+\ell$}
\DashArrowLine(72,12)(8,12){5}  \Text(40,8)[t]{$H^-_i,\ell$}
\Gluon(40,60)(40,90){4}{4} \Text(45,75)[l]{$\uparrow,q,\mu$}
\end{picture}
\end{center}
The CP violating piece of the effective Hamiltonian is parametrized as
$$
{\cal H}^{\Delta S=1}= (G_F/\sqrt{2}) \tilde{C}
( \bar s T^a \gamma_\mu(1+\gamma_5)d )
                       \times
                   \sum_q (\bar q T^a \gamma^\mu q )  \ .
$$
At the electroweak scale, the Wilson coefficient is
$$
\tilde{C}=-\alpha_s
\sum_i{\xi_{di}\xi^*_{si}\over 6\pi}
       {m_W^2\over M_Q^2} F\left({m_{H_i}\over M_Q^2}\right)
\ ,
$$
$$
F(h)=\left[
  { (2h-3)h^2 \ln h\over  (1-h)^4 }
+ { 7-29h+16h^2 \over 6(1-h)^3} \right] \ .
$$
The electromagnetic penguin and long distance effects can  
contribute to the imaginary part of the $\Delta I=3/2$ amplitude and 
give a small contribution.  This is analyzed in detail in Ref.~\cite{bck}.
In our decoupling limit, the result is 
\begin{equation}
\epsilon' / \epsilon
 = -1.9 \times 10^{-5} 
\hbox{Im}  ({{\cal A}_{sd} / (0.058)^2})R_Q^2
\end{equation}
$$
 = \pm 1.9 \times 10^{-5} \
(\sqrt{(156{\cal F})^2 +1} -156{\cal F})^{1\over2} {R_Q /
\sqrt{2}} \ ,
$$
using the constraints in Eq.(\ref{eq:dmineq}).
For $R_Q =1$  and ${\cal F} \approx 0$ (or $- 0.3)$, $\epsilon'/\epsilon =
1.4 \times 10^{-5}$ (or $1.3 \times 10^{-4})$, which is roughly the same
order of magnitude as the KM model.  One can of course makes the model
more superweak~\cite{superweak} by setting the scale higher (and $R_Q$
smaller) and ${\cal A}_{sd}$ larger.  For ${\cal A}_{sd}$ of order one, 
$M_Q$ is roughly 100 TeV.

\section*{Constraints from Strong CP $\theta_{QCD}$\\
and the induced KM phase}

In both RH and LH models, $\theta_{\rm QCD}$ is only induced starting at
the two-loop level, via generation of complex down-flavor quark masses as 
long as the coupling $\kappa$ exists.  A typical diagram is shown in
Fig.~1 in Ref.~\cite{bck}; this effect does not require more than one
flavor of down-quark.  However, it does require both charged Higgs to be
involved.
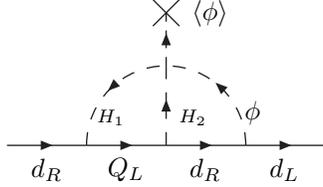
\begin{center}
\begin{picture}(100,75)(0,-20)
\ArrowLine( 0,0)(30,0) \Text(15,-5)[t]{$d_R$}
\ArrowLine(30,0)(60,0) \Text(45,-5)[t]{$Q_L$}
\ArrowLine(60,0)(90,0) \Text(75,-5)[t]{$d_R$}
\ArrowLine(90,0)(120,0) \Text(105,-5)[t]{$d_L$}
\DashCArc(60,0)(30,0,180){5}   \Text(45,10)[rb]{$_{{H}_1}$}
\DashArrowLine(60,0)(60,30){5} \Text(65,10)[lb]{$_{{H}_2}$}
                               \Text(90,10)[lb]{$\phi$}
\ArrowArc(60,0)(30,130,135)
\ArrowArc(60,0)(30,45,50)
\Line(55,45)(65,55) \Line(55,55)(65,45) \Text(70,50)[l]{$\langle\phi\rangle$}
\DashArrowLine(60,30)(60,50){5} 
\end{picture}
\end{center}
Roughly,
$\theta_{\rm QCD}\sim 
g^2 I\ {\rm Im}({\cal A}^{dd}_{12}\ {\cal B}_{12})/(16\pi^2)^2$. 
The factor $I$, of order one, is given by the integral 
$$ I =\int_0^1  {dz   \over (1-z)}
\int_0^{1-z}dx  \ \times $$
$$
\left[
{zm_{H_2}^2+xM_Q^2+ym_{H_1}^2 
\over  
 zm_{H_2}^2+xM_Q^2+ym_{H_1}^2 -z(1-z)M_{\phi^0}^2} 
\ \cdot
\right. $$
$$\left.  \quad \log {z(1-z)M_{H^0}^2 \over  zm_{H_2}^2+xM_Q^2+ym_{H_1}^2} 
      - ( m_{H_1} \leftrightarrow m_{H_2}) 
  \right]  
\ ,
$$
with the Feynman parameters $y=1-x-z$.
The integral $I$ vanishes at the degenerate case
$m_{H_2}=m_{H_1}$, but its size approaches~\cite{bck2} 1 
as $m_{H_2}\to\infty$.  This non-decoupling phenomena is not surprising
because, in the large $m_{H_2}$ limit, the CP is a broken symmetry. 
Numerically, the present constraint, $\theta_{\rm QCD} < 10^{-9}$, can
easily be accommodated.  In addition, there are also three loop diagrams
due to the gluonic contribution.  A typical graph is shown below.
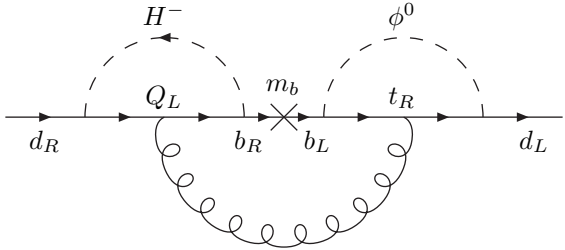
\begin{center}
\begin{picture}(220,100)(0,-55)
\ArrowLine( 0,0)(30,0)  \Text(15,-5)[t]{$d_R$}
\ArrowLine(30,0)(60,0)  \ArrowLine(60,0)(90,0)  
\Text(60,5)[b]{$Q_L$}
\ArrowLine(90,0)(105,0)   \ArrowLine(105,0)(120,0) 
\Line(100,-5)(110,5) \Line(100,5)(110,-5)
\Text(118,-5)[t]{$b_L$}
\Text( 92,-5)[t]{$b_R$}
\Text(105,10)[b]{$m_b$}
\ArrowLine(120,0)(150,0) \ArrowLine(150,0)(180,0)
\Text(150,5)[b]{$t_R$}   \Text(200,-5)[t]{$d_L$}
\ArrowLine(180,0)(210,0)
\DashArrowArc(60,0)(30,0,180){5}   \Text(60,35)[b]{$H^-$}
\DashCArc(150,0)(30,0,180){5}  \Text(150,35)[b]{$\phi^0$}
\GlueArc(105,0)(45,180,360){4}{10}
\end{picture}
\end{center}
This contribution is independent of the coupling $\kappa$.  However,
unless $\kappa$ happens to be very small, it may
not be competitive with the two loop contributions because of the KM angle
suppression and the additional loop factor~\cite{bck2}.  Both 
the 2-loop and the 3-loop contributions also exist
generically in LH models.  However, they are typically numerically smaller
in that case.  The two loop contributions disappear 
(in both RH and LH models) of
course when only one complex scalar boson is used as mentioned earlier.

Since CP is broken at higher energy, a non-vanishing KM phase $\eta$
(defined in the Wolfenstein parametrization)~\cite{eta} can in general be
loop induced.  It can originate either from the loop-induced complex mass
matrix or from that of the complex kinetic energy terms.  The contribution
from complex mass matrix is similar to the analysis of $\theta$ and is
therefore small.  
The contribution from complex kinetic terms is
induced at the one loop level in LH models as analyszed in
Ref.~\cite{gg}.  It is in general also suppressed by some small KM
mixing angles and small mass ratios and therefore numerically tiny
enough to be ignored phenomenologically.  In contrast, in RH models, since
the KM phase is related only to the rotation of the left-handed quarks, 
such contribution will not arise until at the two loop level as given in
the following figure.  Therefore induced $\eta$ is even 
smaller~\cite{bck2}.
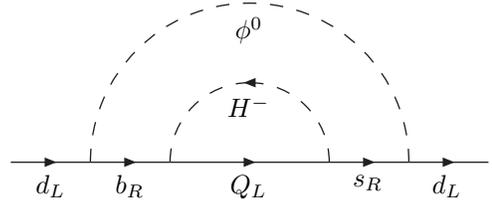
\begin{center}
\begin{picture}(200,75)(0,-10)
\ArrowLine( 0,0)(30,0)  \Text(15,-5)[t]{$d_L$}
\ArrowLine(30,0)(60,0)  \Text(45,-5)[t]{$b_R$}
\ArrowLine(60,0)(120,0)  \Text(90,-5)[t]{$Q_L$}
\ArrowLine(120,0)(150,0) \Text(135,-5)[t]{$s_R$}
\ArrowLine(150,0)(180,0) \Text(165,-5)[t]{$d_L$}
\DashCArc(90,0)(60,0,180){5}       \Text(90,55)[t]{$\phi^0$}
\DashArrowArc(90,0)(30,0,180){5}   \Text(90,25)[t]{$H^-$}
\end{picture}
\end{center}

\section*{Constraints from electric dipole moments}

A down-flavor quark EDM,
however, is generated at the two-loop (or three-loop) level, in parallel
with the generation of complex down quark masses discussed above.  The
typical contribution is given by diagrams similar to those for $\theta$,
except with an external photon attached to internal charged lines. An
estimate of the two loop contribution gives EDM which is consistent with
the current experimental bound.
The contribution from chromo-electric dipole moment of gluon (the Weinberg
operator) won't arise until three loop level (even with $\kappa$ coupling) 
and therefore expected to be small.  The electron couples only indirectly
with the CP violating sector, so its EDM vanishes at two loops
and the three loop contributions are insignificantly small.

\section*{$B^0$--$\bar {B^0}$ Mixing, $b \to s \gamma$ and Other
Constraints}

Another (much weaker) constraint to be considered is that from the
$B^0_{s,d}$ mass splitting~\cite{bck}. 
Using the usual estimate of strong form factors involved and
the experimental value for $\Delta M_{B^0}$, we have 
$$\delta(\Delta
M_{B^0}) /\Delta M_{B^0} = 1.1\times 10^{-3}\, R_Q^2 \,
\hbox{Re}\left({\cal A}_{bd} /0.058^2\right)^2 \ , $$
so even taking ${\cal A}_{bd}= (0.15)^2$, the fractional contribution
is only about $5\%$ for $M_Q=300$ GeV.  

In RH models, the operator induced by the exotic sector has
helicity opposite to the SM contribution, the two do not
interfere in the rate.
In the decoupling limit
with ${\cal F}_{b\to s\gamma} 
\equiv \delta B(b\rightarrow s\gamma)/B(b\rightarrow s\gamma)_{\rm SM}$,
we have
$$
{\cal F}_{b\to s\gamma} 
=6.4\times  10^{-6}
\left| R_Q^2 \cdot  {0.0389\over V_{tb}V^*_{ts}}
\cdot  {{\cal A}_{bs}\over  (0.058)^2}\right|^2 \ .
$$
Furthermore, the relevant parameter ${\cal A}_{bd}$ is not subjected
to constraints from $\epsilon$ or $\epsilon'$. If it is
of the same size as ${\cal A}_{sd}$, the deviation from the SM would
be negligible and the future $B$ factory 
would observe only a collpased unitarity KM angle~\cite{Mele}. However,
if ${\cal A}_{bd} \gg {\cal A}_{sd}$, the triangle can looks
substantially different from that predicted by standard KM model. It
is worthwhile to point out that, in LH models, the exotic contribution
gives rise to operator that will interefere with that of SM and therefore
the model is more severely constrained~\cite{bck2}.  

\section*{Decays of New Particles}
In the generic RH model, $h$ and $Q$ can be assigned a new
conserved quantum number which guarantees a stable lightest exotic
particle, either $H_1$ or $Q$. 
However, when $q_h =-1$, an interaction, $h_\alpha L_i L_j$, is allowed,
which can lead to $H^-$ (on-shell or off-shell) decays into $l^-\nu$.
Even so, lepton number is still conserved, just as in the
SM, since $Q$ and $H$ will naturally carry  lepton number ($L=\pm
2)$.
Another way for $H$ to decay is to introduce a second
Higgs doublet and let $H$ couple  to two different Higgs doublets.  In
that case $H$ can decay into a neutral Higgs, plus a charged Higgs which in
turn decays into ordinary quarks and leptons.

\section*{Spontaneous broken CP symmetry}

To show how the above softly broken CP symmetry can in fact
originate from a
spontaneously broken one, one can first add a CP-odd scalar,
$a$, which develops a non-zero vacuum expectation value (VEV) and
breaks CP.  However, this scalar will in general couple to
$\bar{Q_L}Q_R$ and give rise a complex tree level $M_Q$ and,
therefore, a tree level $\theta_{\rm QCD}$.  To avoid this, one can
add another scalar singlet, $s$, which is CP-even and impose a discrete
symmetry which changes the signs of both $a$ and $s$ and nothing else.  
As a result, a term such as $ i a \bar Q \gamma_5 Q$ is forbidden.  
The only additional term relevant for CP violation is $i\left[ s \,a
\,({h_1}^{\dag} h_2 - {h_2}^{\dag} h_1) \right]$ which generates 
a complex $(m^2)_{12}$ after both $s$ and $a$ develop VEVs.
The extra neutral Higgs bosons will  mix with the
SM Higgs, but since $s$ and $a$ do  not couple to fermions
directly, they have tiny scalar-pseudoscalar coupling to
fermions  only at loop level.  As a result, 
the CP phenomenology considered below applies equally
well to both the softly and spontaneously broken versions of our model.

\section*{Interplay between Strong and Weak CP Phase}

The RH models provide a good example to look at the 
subtlety, raised in Ref.~\cite{interplay}, involving the interplay of 
the CP phases between the strong and the weak interactions. 
For our purposes, we can focus on the reduced effective theory which
contains CP-conserving Standard Model-type interactions and vanishing
$\theta_{QCD}$, plus the new, induced superweak interaction
with the strength $C^R_{\Delta S=2}$ defined in Eq.~(4).
We shall consider, for the sake of argument, the scenario in which  
the up quark is massive while $m_d$ is zero.
Without the new  $C^R_{\Delta S=2}$ 
interaction the parameter $\theta_{QCD}$ is then
unphysical, with CP a good symmetry.
(By the usual argument, with a
massless quark present --- in this case, the $d$ quark --- the
right-handed component of that quark can be rotated to absorb
$\theta_{QCD}$ via the axial anomaly, while otherwise leaving the
lagrangian invariant.) 
With the addition of the induced interaction 
${\cal H}^{\Delta S=2}$, however, $\theta_{QCD}$ becomes  physical, as
can be seen by considering the following cases:

(a) If $C^R_{\Delta S=2}$ 
is complex and $\theta_{QCD}=0$, CP is violated. In this
case, the correct (non-zero) value for $\epsilon$ can be calculated
without complication;  all hadronic matrix elements (modulo absorptive
contributions) can correctly be assumed to be real.  It is illuminating
to consider the calculation of $\epsilon$ in another basis, which is
obtained
by a phase rotation of $d_R$ such that $C^R_{\Delta S=2}$ becomes real and
$\theta_{QCD}$
non-zero.  Since the two theories are the same, one must arrive at the
same result for $\epsilon$.  One can thus draw 
a rather surprising conclusion: $\theta_{QCD}$
 {\em can also, in certain situations, contribute to $\epsilon$}.

In fact, from the way we obtain the $\theta_{QCD}$ contribution to
$\epsilon$ in this example, one realizes that there is an important
subtlety here. {\it The actual contribution from $\theta_{QCD}$ to
$\epsilon$ is correlated to the explicit mechanism of CP violation},
which in our current example is the superweak $C^R_{\Delta S=2}$. 
A related result is that  when $\theta_{QCD}$ is not zero, how each
hadronic
matrix element develops a phase {\em also depends on the particular
electroweak
mechanism of CP violation in the theory}.  In the present case, the CP 
violating coupling also happens to be the chiral symmetry 
breaking phase.

Another lesson one learns is that the usual argument 
which concludes that the contribution of
$\theta_{QCD}$ to CP-violating quantities such as the neutron electric
dipole moment (edm) must
be proportional to $m_u m_d$, is not strictly correct, a counterexample is
offered by the simplified model presented above.  The role of $m_d$ is
replaced by the coupling $C^R_{\Delta S=2}$.  Of course, $C^R_{\Delta
S=2}$
breaks the chiral symmetry associated with $d$ quark, so that the $d$
quark will certainly pick up mass at some (probably higher-loop) level,
but the point is that the $C^R_{\Delta S=2}$ coupling plays a much more
direct role in the contribution of
$\theta_{QCD}$ than even the induced $m_d$!

Now we come to an apparent paradox whose resolution gives even further
insight into the interplay between strong and weak CP phases.

We parenthetically noted above that if redefinition of the quark phases
generates an imaginary part for the quark mass matrix {\em not
proportional} to the identity matrix, the low-energy meson states must be
suitably reinterpreted to ensure stability of the vacuum around which we
carry out perturbation theory. This redefinition explicitly reintroduces
the phase(s) rotated from the couplings into certain hadronic matrix
elements, to ensure rephasing invariance. If both $m_d, m_u$ vanish, then
arbitrary rotation of the corresponding right-handed quarks seems to have
no effect on vacuum stability, since the mass matrix is left real and
diagonal (only $m_s$ non-zero). Then, apparently, all phases may be
arbitrarily rotated away, and with them, any possibility of CP violation.
Specifically, consider the following variant of the two cases already
considered:

(b) Let $C^R_{\Delta S=2}$ be complex, but take $\theta_{QCD}$ to be zero. If {\em
both}
$m_u, m_d$ are strictly zero, is CP conserved or violated? 
At first glance, one might  claim the phase in $C^R_{\Delta S=2}$ to be unphysical,
since a
combined phase rotation of the form $u_R\rightarrow e^{-i\delta}u_R$ and
$d_R\rightarrow e^{i\delta}d_R$ can make $C^R_{\Delta S=2}$ real and
maintain $\theta_{QCD}=0$.  It is very tempting to claim  that CP
violation is proportional to $m_u$  for a small up quark mass and further
that there is no CP violation when $m_u\rightarrow 0$ because the phase of 
$C^R_{\Delta S=2}$ then becomes absorbed.

This conclusion is {\em incorrect}, however,  because we have ignored the
vacuum
degeneracy in the case of massless $u$ and $d$ quarks. Different
choices of vacua would give different CP violation.  It is true that
there exists one very special vacuum where CP is conserved. However, a
general vacuum posseses chiral condensate with a phase uncorrelated to
that of $C^R_{\Delta S=2}$, and thus CP violation usually occurs, {\em even if} $C^R_{\Delta S=2}$
is real (since what is important is the relative phase between the vacuum
and $C^R_{\Delta S=2}$). This idea can be
demonstrated directly in the chiral effective lagrangian approach.
The chiral field $\Sigma$ (3$\times$3 unitary matrix) can be perturbed
around a vacuum configuration diag$(e^{-i\phi},e^{i\phi},1)$. If
$C^R_{\Delta S=2}$ is turned off, the strong interaction is independent of $\phi$
because of the chiral symmetry.  However, with $C^R_{\Delta S=2}$, the phase $\phi$
has physical meaning and has implications with respect to CP violation.
Now we include effects of the real quark masses $m_u\ne 0$ and $m_d\ne 0$.
Their net effect is simply to pick out a particular vacuum,
with   $\Sigma=$diag$(1,1,1)$. In this case of vacuum alignment, a complex
$C^R_{\Delta S=2}$ is necessary, but also sufficient, for CP violation, since again it
is the relative phase between $C^R_{\Delta S=2}$ and the vacuum that is important. In
some sense, the (possibly infinitesimal) up and down quark masses {\em
enforce} CP violation, in the particular case that $C^R_{\Delta S=2}$ is real, whereas
in the massless case, CP violation is still generally expected via the
vacuum phase.

\section*{Conclusion}

\noindent
We have reviewed a new (RH) class of models whose CP violation is solely
mediated by exotic Higgs bosons and fermions that couple to the right
handed quarks, and compared with another (LH) class of models whose exotic
particles couple to the left-handed quarks.  The model naturally prevents
the strong CP problem.  Both classes of models are
surprisingly similar to the KM model in the sense that the CP-breaking
mechanism is seemingly milliweak(if the exotic particle scale is chosen to
be as low as possible), while its phenomenology (as studied
here) is quite
superweak-like. The phenomenological distinction between the two will
likely be made clear in experiments planned for the B factory.
A careful and detailed analysis of such issues is
clearly necessary and is in progress~\cite{bck2}.

D.~B.-C. and W.-Y.~K. are supported by a grant from the DOE of USA, and
D.~C. by the NSC of R.O.C.
We thank
H. Georgi,
S. Glashow,
P. Frampton,
R. Mohapatra,
and
L. Wolfenstein
for very useful discussions.

\def\baselinestretch{1.2}

\end{document}